\documentclass[pre, reprint,superscriptaddress, showkeys]{revtex4-2}

\usepackage[utf8]{inputenc}
\usepackage[english]{babel}
\usepackage{graphicx}
\usepackage[dvipsnames]{xcolor}
\usepackage{hyperref}
\usepackage{amsmath}
\usepackage[
  separate-uncertainty = true,
  multi-part-units = repeat
]{siunitx}
\usepackage[version=3]{mhchem} 

\begin{document}

\title{Immersed Cantilever Apparatus for Mechanics and Microscopy}
\author{Akash Singh}
\affiliation{Université de Lyon, Université Claude Bernard Lyon 1, CNRS, Institut Lumière Matière, F-69622, VILLEURBANNE, France}
\author{Michio Tateno}
\affiliation{Research Center for Advanced Science and Technology, University of Tokyo, Japan}
\author{Gilles Simon}
\author{Loïc Vanel}
\author{Mathieu Leocmach}
\affiliation{Université de Lyon, Université Claude Bernard Lyon 1, CNRS, Institut Lumière Matière, F-69622, VILLEURBANNE, France}
\email{mathieu.leocmach@univ-lyon1.fr}

\begin{abstract}
    We present here a novel cantilever based apparatus to perform translational stress or strain controlled rheology in very soft solids, and obtain simultaneous confocal imaging of the 3 dimensional microstructure. The stress is measured using eddy based sensors. Both the stress and strain are controlled by applying PID control loops on measured quantities and changing position using a micromanipulator. To get rid of surface tension forces, the sample and cantilever are immersed. This enables stress measurement and control down to $\SI{3}{\milli\pascal}$. With this apparatus, we can independently apply shear and normal stress, or strain, with same precision. We demonstrate the technical capability of the setup with steady shear strain or stress experiments on a soft protein gel system. The simultaneous confocal imaging offers insight into the macroscopic breaking observed in an increasing shear strain experiment
\end{abstract}

\maketitle


\section{Introduction}
\label{sec:introduction}
Soft solids such as gels, foams, and fiber tissues, have self-assembled microstructure, and their multiscale mechanical responses emerge as wide variety of rheological and fracture behaviors. To understand these phenomena, three dimensional (3D) visualization of microscopic structural changes against mechanical stimulus is desirable. However, since elasticity scales with the typical microstructure size $\xi$ as $k_\mathrm{B}T/\xi^3$, ($k_\mathrm{B}$: Boltzmann constant, $T$: temperature), soft solids with microstructure observable by optical microscopy have usually very small elastic modulus and are too soft to be stressed in a controlled way. In the present paper, we will showcase an apparatus that allows the application or measurement of extremely low stresses on soft materials in both shear and normal direction, while observing the 3D microstructure by optical (confocal) microscopy.

Coupling a confocal microscope with a commercial or custom rotational rheometer seems to be the most straightforward way to observe in real space the microstruture evolution upon mechanical stimulation. This solution is well adapted to study samples under a steady shear rate~\cite{derksConfocalMicroscopyColloidal2004, ballestaSlipFlowHardSphere2008,rajaramMicrostructuralResponseDilute2010a, duttaDevelopmentConfocalRheometer2013a,arevalo2015stress}, oscillatory shear~\cite{ smithYieldingCrystallizationColloidal2007a} or constant stress~\cite{paredesShearBandingThixotropic2011,lindstromStructuresStressesFluctuations2012,chanSimpleShearCell2013, sentjabrskajaCreepFlowGlasses2015}. The cone-plate geometry is often chosen in rotational rheometers in order to achieve homogeneous shear~\cite{derksConfocalMicroscopyColloidal2004,besselingQuantitativeImagingColloidal2009,rajaramMicrostructuralResponseDilute2010a,paredesShearBandingThixotropic2011,chanSimpleShearCell2013}. However in such geometry the only way to observe the whole thickness of the gap and thus quantify the effect of wall slip is to observe very close to the axis of the (truncated) cone, where the shear is actually not homogeneous~\cite{Ballesta2013}.

Translational shear cells with plate-plate geometry offer homogeneous shear and are better suited to integrate with optical microscopy due to their simpler and less expensive design. Thus, they have been widely used to study yielding transition in soft solids~\cite{cohenSlipYieldBands2006a,smithYieldingCrystallizationColloidal2007a, besselingQuantitativeImagingColloidal2009,hsiaoRoleIsostaticityLoadbearing2012,Chikkadi2012,koumakisYieldingHardSphereGlasses2012,boitteObservationWheatFlour2013,knowltonMicroscopicViewYielding2014,Lin2014,aimeStresscontrolledShearCell2016}. However shear cells usually have small plate surface area to achieve a high degree of parallelism, leading to a small and thus difficult to measure net force. Indeed, most translational shear cells lack stress measurement. Only few works~\cite{Lin2014,aimeStresscontrolledShearCell2016} have explored the possibility to have stress measurement and, to our knowledge, stress control is only available in the setup proposed in Ref.~\cite{aimeStresscontrolledShearCell2016}. However this control is less sensitive than in rotational stress-controlled rheometer, restricting its usage to rather large stresses ($>\SI{1}{\pascal}$). Furthermore, normal stress cannot be controlled and is never measured.

To be observed with confocal microscopy, the microstructure needs to be at least a micrometer large, leading to an extremely soft material. For instance, a colloidal gel made of micron-size particles with $\SI{10}{\micro\metre}$ structural pore size will have moduli of the order of \SI{10}{\milli\pascal} and a yield stress closer to \SI{1}{\milli\pascal}~\cite{tsurusawaDirectLinkMechanical2019}. Such stresses are too low to be reliably applied by most commercial rheometers, and even less so by shear cells. That is why most rheo-confocal studies on colloidal gels have been performed by controlling the strain or the strain rate, with no measure of the stress response~\cite{smithYieldingCrystallizationColloidal2007a,lindstromStructuresStressesFluctuations2012,hsiaoRoleIsostaticityLoadbearing2012}. Indeed, the stress response is often extrapolated from quantitative measurements done on similar systems with much smaller building blocks~\cite{Pham2006,lindstromStructuresStressesFluctuations2012}.

Cantilever deflection is another major approach to measure mechanical properties. Since Galileo~\cite{galileiDiscorsiDimostrazioniMatematiche1638}, its principle has been used to quantify material properties, from geological~\cite{bransbySoilDeformationsCantilever1975} to atomic scale. It is at the basis of atomic-force microscopy~\cite{binnigAtomicForceMicroscope1986}, surface-force apparatus~\cite{taborDirectMeasurementNormal1969,israelachviliMeasurementForcesTwo1978}, and several biosensors~\cite{fritzCantileverBiosensors2008}. The deflection of the cantilever is often measured by the reflection of a laser on its tip~\cite{meyerNovelOpticalApproach1988a}. For centimetric cantilevers in possibly turbid environments, eddy current sensors offer a good trade off between precision and compactness. Compared to capacitive sensors, eddy current sensors offer a larger dynamic range and are unperturbed by changes in medium conductivity. A compression and stretching device based on a decimeter-long cantilever blade for which deflection was measured by an eddy current sensor has successfully quantified the viscosity of cell aggregates~\cite{marmottantRoleFluctuationsStress2009, stirbatFineTuningTissues2013}, and the surface tension of liquids, biological tissues and yield-stress fluids~\cite{Mgharbel2009,Geraud2014,Jorgensen2015}. Furthermore, this apparatus has been coupled to observation of the microstructure (although, not in 3D)~\cite{Mgharbel2009,marmottantRoleFluctuationsStress2009}. However, since the sample was here in contact with air, the theoretical sensitivity of \SI{0.1}{\milli\pascal} on bulk stress was never reached as surface tension forces were dominating. Thus, even if the  concept of using cantilever as force measurement apparatus is quite old, its usage for shear rheology of yield stress solids combined with confocal microscopy observation, has not been reported, to our best knowledge.

\begin{figure}[t]
\centering
\includegraphics[width=\columnwidth]{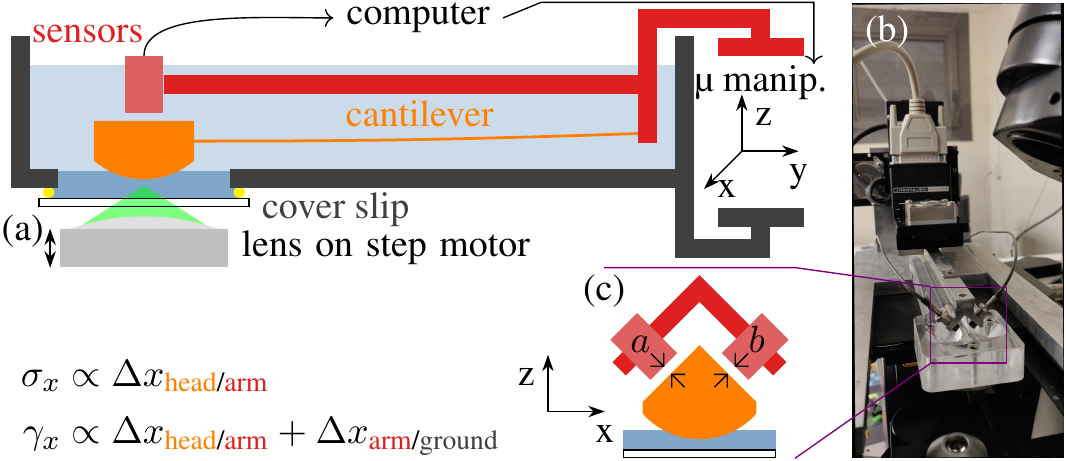}
\caption{Design of the ICAMM with (a) a schematic side section view, (b) an angled front view of the actual device and (c) a schematic front section view. The distances $a$ and $b$ measured by the sensors are highlighted by pairs of facing arrows. In the schematics, elements are coloured with respect to their reference frame: orange for the head and the cantilever, red for the sensors and the arm, dark gray for the tank (ground frame). The gel sample is shown in blue, whereas the liquid permeating it and surrounding the head is shown in light blue.%
\label{fig:apparatus}}
\end{figure}

Improving on the principle of the above compression and stretching device, we have developed a novel apparatus that can perform both shear and normal tests while capturing 3D microstructure by confocal microscopy. This device, named ``Immersed Cantilever Apparatus for Mechanics and Microscopy'' (ICAMM), is sketched in Fig~\ref{fig:apparatus}. The setup offers stress and strain measurement, and can apply controlled stress or strain independently in shear and normal direction using PID loops. The sensitivity of this setup is not limited by interfacial forces, and is the same in both directions, which is another advantage over other reported methods. In the paper, we elaborate on our set-up design in Section~\ref{sec:working_principle}. The complication with this setup arises with the selection of the cantilever and chemical composition of the soft system and surrounding buffer. Section~\ref{sec:cantilever_calibration} covers the selection of the cantilever and the required calibrations.  In Section~\ref{sec:test_exp}, we test this set-up using casein gel, covering the chemical preparation, testing of control loops, and demonstration of controlled shear stress and strain experiment. Finally, we conclude with a summary of the observations and other potential uses for this setup. 

\section{Working Principle}
\label{sec:working_principle}
The apparatus design and main components are shown in Fig.~\ref{fig:apparatus}. The working principle is based on measuring the deflection of a cantilever to obtain the stress applied to the sample. At the end of the cantilever, a rigid `head' (orange in Fig.~\ref{fig:apparatus}) has its bottom in contact with the top of the sample. Two metallic targets of \SI{8}{\milli\metre} width and \SI{0.3}{\milli\metre} thickness are glued at \SI{45}{\degree} on both sides of the head ( Fig.~\ref{fig:apparatus}c). The distance $a$ and $b$ to each of these targets is measured by an eddy current position sensor (EPS08-C3.5-A/M, Micro-Epsilon, light red on Fig.~\ref{fig:apparatus}) respectively in ferromagnetic or non ferromagnetic target mode. Indeed, using different modes and targets (stainless steel and aluminum respectively) is necessary to avoid interference between the sensors. Both sensors are mounted on a rigid ($\SI{10}{\kilo\newton/\metre}$) `arm' (red on Fig.~\ref{fig:apparatus}) to which the base of the cantilever is also clamped. Sensor readings thus give direct access to the position of the head of the cantilever with respect to its base, that is to say the deflection of the cantilever.

The arm is mounted on a micromanipulator (MP285, Sutter Instrument) allowing three axis translation with respect to the ground frame via step motor (16 steps per \si{\micro\metre}). Thus, the position of the head with respect to the ground frame (e.g. $x_\mathrm{head/ground}$) is obtained by summing the displacement of the micromanipulator (e.g. $x_\mathrm{arm/ground}$) with the displacement of the head obtained from the sensors (e.g. $x_\mathrm{head/arm}$). Since the bottom of the sample is fixed with respect to the ground frame, the position of the head with respect to the ground frame can be converted to a macroscopic strain field, knowing the geometry.

The deflection of the cantilever can be converted to a force. However, further conversion to a stress field in the volume of the sample is in general made more complicated by the contribution of interfacial forces acting between cantilever, sample and air~\cite{Jorgensen2015}. Since we are dealing with gel samples permeated by a solvent, we are able to get rid of surface tension effects by fully immersing the sample (dark blue on Fig.~\ref{fig:apparatus}a), the cantilever and its head into the same solvent (light blue on Fig.~\ref{fig:apparatus}a). Provided a fine tuning of the solutes in this solvent (see Section~\ref{sec:sample_prep}), the gel network can maintain its mechanical properties while immersed. A collateral benefit of the immersion is a buoyancy force acting on the head, that partially counteracts its weights, providing the opportunity to use a softer cantilever without experiencing its plastic bending.

The solvent is contained by a machined PMMA tank (dark gray in Fig.~\ref{fig:apparatus}). The bottom of the tank (\SI{2}{\milli\metre} thick) has a circular ($\SI{15}{mm}$ diameter) hole to allow observation with an inverted optical microscope. This hole is reversibly mounted and sealed (Teflon tape $>\SI{0.1}{\milli\metre}$ thickness) with a glass coverslip (\SI{30}{\milli\metre} diameter, \SI{0.17}{\milli\metre} thickness) pressed by an inverted conical stainless steel mount piece attached to the tank by three screws. The gel sample is sandwiched between this coverslip and the head of the cantilever, as sketched in Fig.~\ref{fig:apparatus}a. The whole apparatus can be used either alone for purely mechanical measurements, or mounted on an inverted microscope. The micromanipulator and the tank are connected to a rigid stainless steel base that can be screwed to a standard XY microscope stage, here the motorized stage of a Leica SP5 confocal microscope. The whole apparatus weighs approximately \SI{3}{\kilo\gram}. Mounting and unmouting from the stage can be done in a few minutes.

\begin{figure}
    \centering
    \includegraphics[width=3.325in]{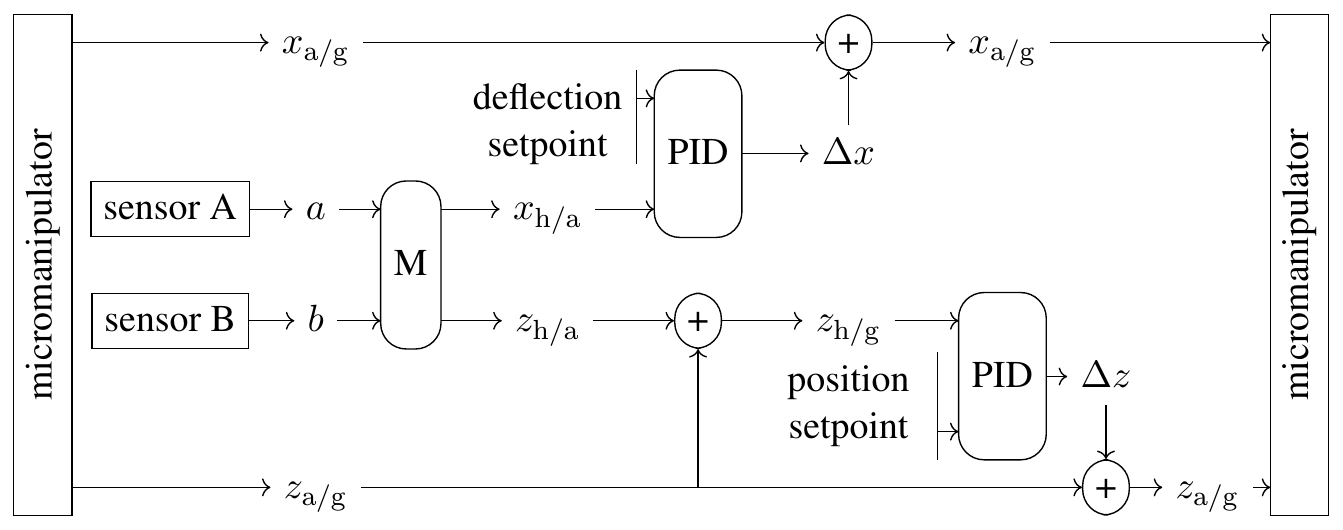}
    \caption{Diagram of the control loops in the case of a constant shear stress and a constant normal position. For the sake of space, names of reference frames are shortened to their initials.}
    \label{fig:pid}
\end{figure}

The output of each sensors is read and digitized by a DT3100-SM (Micro-Epsilon) electronics. Digital readings from the sensors (ethernet) and the micromanipulator (serial) are centralized on the host PC by a Python program that also actuates the micromanipulator. Source code of the program can be found at~\footnote{\url{https://doi.org/10.5281/zenodo.4892987}}. Using PID control loops enables either stress or strain control on each axis, as shown on Fig.~\ref{fig:pid}. If the displacement required is less than a few tens of microns, the loop performs at \SI{9}{\hertz}, mostly limited by the response time of the serial communication with the micromanipulator.

\section{Choice and calibration of the cantilever}
\label{sec:cantilever_calibration}

The choice of the cantilever is crucial in the current apparatus. In order to obtain the same stiffness in every direction of flexion, we settled to a circular section. This sets the deflection in response to a force $F$ on the head to
\begin{equation}
\delta = \frac{64 L^3}{3\pi E D^4} F,
\label{eq:defl}
\end{equation}
where $L$ and $D$ are the length and diameter of the cantilever and $E$ its young modulus.
Aside from flexion, a circular cantilever can display torsion that may disturb our measurements. For a force $F$ applied tangentially to the bottom of the head (at a distance $\ell_b$ from the axis of the cantilever), the sensors placed at distance $\ell_s$ from the axis of the cantilever will measure a displacement due to torsion
\begin{equation}
    \delta_T = \frac{64 \ell_b \ell_s L}{\pi D^4}\frac{1+\nu}{E}F,
\end{equation}
where $\nu$ is the Poisson ratio of the material. We thus have $\frac{\delta}{\delta_T} = \frac{1}{3(1+\nu)}\frac{L^2}{\ell_b \ell_s} \approx 500$ for $L=\SI{20}{\centi\metre}$, $\ell_b=\SI{2}{\milli\metre}$,  $\ell_s=\SI{10}{\milli\metre}$ and $\nu=0.3$. Therefore, the torsion mode is negligible in our measurements but could be an issue for shorter cantilevers.

We have tried cantilevers in pure copper and stainless steel, however they showed too narrow elastic domain for our purpose. We finally settled to copper-beryllium alloy (Cu 98 $\%$ and Be 2 $\%$, GoodFellow CU075340, $\nu=0.3$ and $E= \SI{120}{\giga\pascal}$-$\SI{160}{\giga\pascal}$) for its large elastic domain. In the following, we further characterize a cantilever of length $L \approx \SI{20}{\centi\metre}$ and diameter $D = \SI{1.0}{\milli\metre}$.

\subsection{Geometric calibration}
\begin{figure}
\centering
\includegraphics[width=\columnwidth]{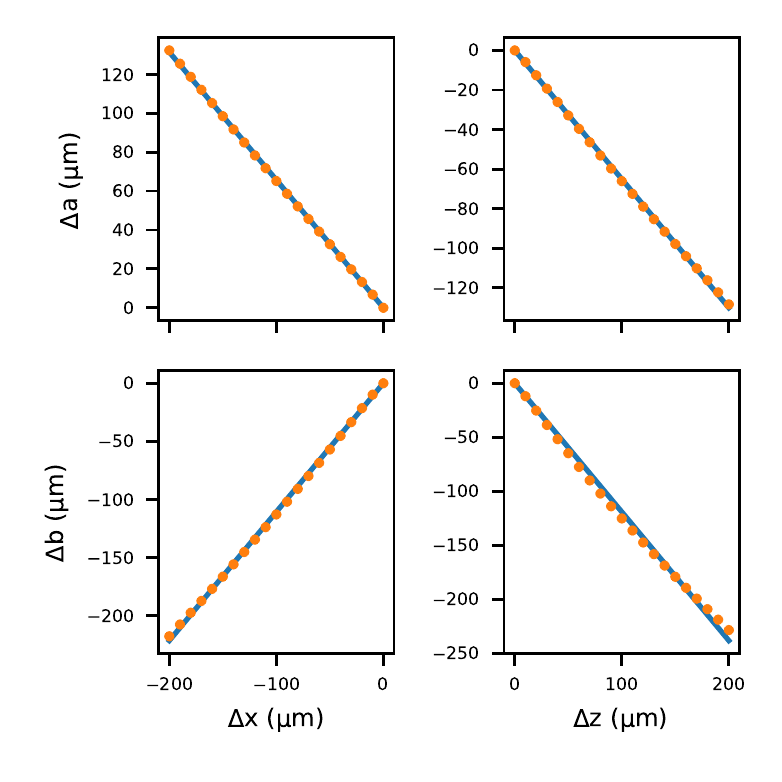}
\caption{A typical geometric calibration of the sensor distances from the head $(a,b)$ to the lab frame of reference $(x, z)$(scattered points). The error in both x and y direction are smaller than the plotted points. A linear fit (line) and matrix inversion gives the value of the calibration matrix $M$ as $\left(\begin{smallmatrix}-0.793&-0.433\\0.736&-0.437\end{smallmatrix}\right)$.}
\label{fig:geometric_calibration}
\end{figure}

On each mounting of the cantilever or the sensors, we perform a geometric calibration so that the reading of the sensors $(a,b)$ is properly converted to the $(x,z)$ coordinate system. The displacement of the cantilever head along the arm ($y$ direction) is negligible since the cantilever length ($L=\SI{20}{\centi\metre}$) is much larger than typical movement of cantilever head ($\lesssim \SI{100}{\micro\metre}$). We physically block the head against an obstacle normal to $x$, make the micromanipulator move by a known distance along $x$ and take the sensor readings 10 times, averaging them to record $(a,b)$. We repeat this procedure every \SI{10}{\micro\metre} up to \SI{200}{\micro\metre} and perform the same along $z$. A typical set of results is shown in Fig.~\ref{fig:geometric_calibration} as scattered data. The error in this plot depends on the repeatability of our measurement. For the sensors the specified repeatability is $<\SI{0.5}{\micro\metre}$. 

If we assume that the cantilever behaves linearly, then the displacements $\Delta x_\mathrm{head/arm}$ and $\Delta z_\mathrm{head/arm}$ should be given as a linear combination of both $\Delta a$ and $\Delta b$. This can be represented by the matrix multiplication:
\begin{equation}
    \begin{bmatrix} \Delta x_{head/arm} \\ \Delta z_{head/arm}  \end{bmatrix} = M \begin{bmatrix} \Delta a \\ \Delta b  \end{bmatrix},
\end{equation}
where $M$ is a $(2\times 2)$ matrix. The four coefficients of $M^{-1}$ are obtained from a linear least square fit of data shown in Fig.~\ref{fig:geometric_calibration}. We can then inverse the matrix to get $M$. The geometric coefficients are of order 0.4-0.7, whereas their uncertainties are of the order of $10^{-5}$. We can thus consider that the relative uncertainty added by the referential change is of order \num{5e-4}.


\subsection{Stiffness calibration ($k$)}

\begin{figure}[t]
\centering
\includegraphics[width= \columnwidth]{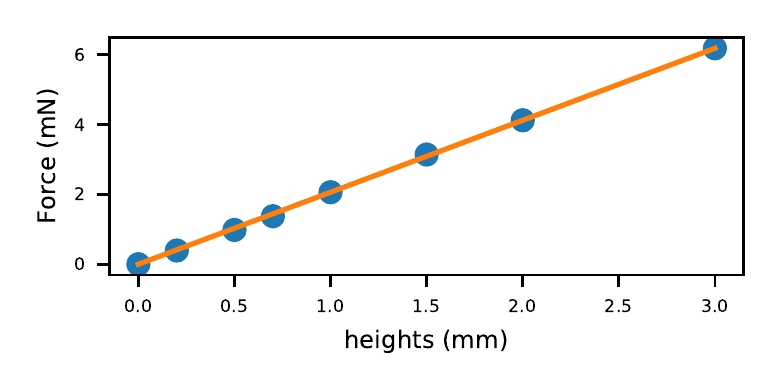}
\caption{Cantilever stiffness calibration. The error in weight are quiet small compare to the actual measurements and hence not visible in the plot. The stiffness coefficient obtained from the linear fit is  $2.059 \pm 0.009 N/m$.}
\label{fig:stiffness_calibration}
\end{figure}

Provided the large dynamic range of eddy current sensors (from \SI{40}{\nano\metre} to \SI{800}{\micro\metre}), forces four orders of magnitude larger than the resolution can be reliably applied. Therefore in our case, a scale precise to \SI{0.01}{\gram} (Denver Instrument, MXX-612) is enough to calibrate the stiffness of the cantilever. 
We start from a position where the head is just touching the scale plate and tare the scale. Then, we lower down the micro-manipulator by a known height, which gives the deflection of the cantilever, while the force is read from the scale. The linearity of the reading is shown in Fig.~\ref{fig:stiffness_calibration} and persists as long as the sensors are not physically touching the head. From a linear fit, we obtain the stiffness coefficient, typically $k=\SI{2.059 \pm 0.009}{\newton/\metre}$. The value matches with the theoretically expected one from (Eq.~\ref{eq:defl}) if $E$ is assumed to be $\SI{130}{\giga\pascal}$, which is within the range of the specification. Furthermore, the cantilever stiffness is at least three orders of magnitude lower than the stiffness of the scale - measured independently to \SI{10}{\kilo\newton/\metre} - which validates the calibration method.

\subsection{From force to stress}
\label{sec:area}

In order to avoid parallelism issues, the part of the head of the cantilever in contact with the gel is a spherical cap of radius of curvature $R_0 = \SI{20}{mm}$, with a base of radius $r_{c}=\SI{6}{mm}$. Between the bottom of the head and the cover slip, we thus have a sphere-plane geometry of minimum gap $h_0$
, with $h_0$ typically $\SI{0.1}{\milli\metre}$. 

Confocal observations will be centered on the vertical axis of the head, with a size of the field of view similar to $h_0\ll R_0$. Therefore, within the field of view, the stress can be considered locally uniform. However, to link the force measured by the cantilever to the stress in the field of view, it is convenient to consider the effective area of a plane-plane geometry of gap $h_0$, exerting homogeneously the stress applied at the lowest point of the head, so that
\begin{equation}
    F = A_\mathrm{eff}\sigma(r=0),
\end{equation}
where $r$ is the distance from the vertical axis of the head, and $A_\mathrm{eff}$ the area of this effective plane-plane geometry.

Without loss of generality, we consider an elastic medium of shear modulus $G$ and a small translation of the head $\delta x$ in the shear direction. This leads to a strain distribution $\gamma(r) = \delta x/h(r)$, and thus a stress distribution on the head $\sigma(r)= G\gamma(r)$. Integrating and equating the forces both in sphere-plane ($h(r) \approx h_0 + r^2/(2R_0)$) and in effective plane-plane ($h(r) = h_0$) geometries, we find
\begin{equation}
    A_{\mathrm{eff}} \approx 2\pi R_0 h_0\ln\left(1+\frac{r_c^2}{2R_0 h_0}\right)
\end{equation}
For our geometrical parameters, $A_\mathrm{eff}\approx\SI{29}{mm^2}$.

\subsection{Systematic and relative uncertainties}
\label{system_uncertainity}
Systematic uncertainties on the stress come from the respective calibrations of $M$, $k$ and $A_\mathrm{eff}$ and sum up to about 10\% uncertainties on the absolute magnitude of the stress measurable with the present apparatus. However, relative uncertainties between two measurements done with the same set of calibrations stem linearly from the resolution of the sensors, $\delta a = \delta b = \SI{40}{\nano\metre}$ for a static measurement. The resolution in stress is thus $\delta \sigma = k \delta a / A_\mathrm{eff} \approx \SI{3}{\milli\pascal}$. 
This is similar to catalog specifications of commercial stress-controlled rheometers (e.g. Anton-Paar MCR 502 with a $R=\SI{25}{\milli\metre}$ cone-plate) and at least an order of magnitude better than published shear-cells~\cite{Lin2014,aimeStresscontrolledShearCell2016}. Furthermore, our apparatus has an equivalent resolution in normal stress, whereas rheometers more often have normal stress resolutions in the range of \SI{1}{\pascal} and shear-cells are to our knowledge not able to measure or to control normal stress.
Also, from Eq.~(\ref{eq:defl}), we deduce that the  stiffness coefficient of the cantilever scales with the diameter and length as $ k \propto D^4 L^{-3}$. In principle, we can bring down the precision to order of \SI{1}{\micro\pascal} using thinner and longer cantilever. This can be useful to study the sub-critical stress behavior in soft colloidal gels.

\section{Test experiments}
\label{sec:test_exp}
\subsection{Sample and surrounding solution}
\label{sec:sample_prep}
\begin{figure}
\centering
\includegraphics[width= \columnwidth]{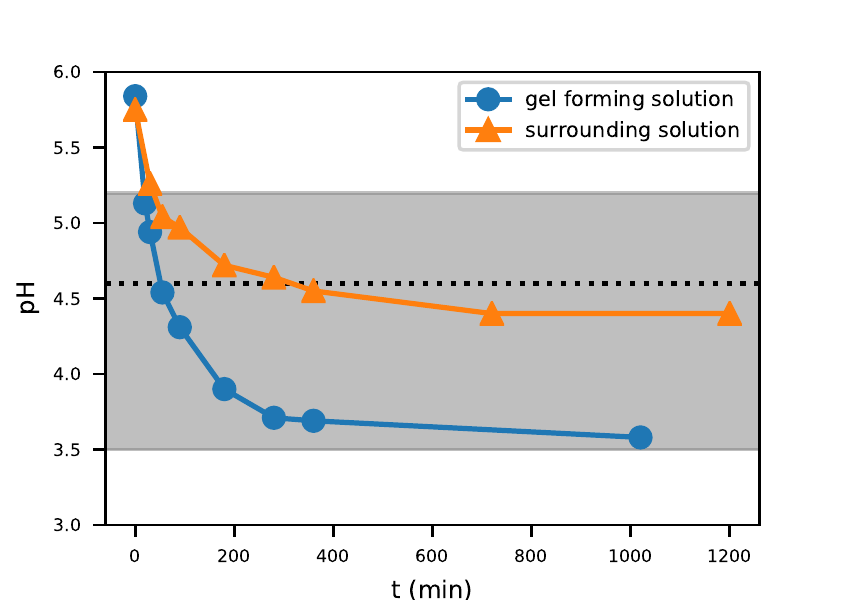}
\caption{Evolution of pH of the surrounding solution and sample solution when not in contact with each other. The zone of pH where casein aggregation is observed is shown in gray~\cite{Leocmach2015}. The horizontal dotted line is the isoelectic pH.}
\label{fig:pH_time}
\end{figure}

We use acid-induced casein gels as test samples. The mechanical behaviour of this soft solid is well characterized, with a linear viscoelastic response up to $\gamma\approx 10~\%$, a weak power-law dependence on the frequency, and irreversible brittle fracture at larger strains~\cite{Roefs1990, Lucey1998, Leocmach2014,keshavarzNonlinearViscoelasticityGeneralized2017}. The mesh is a few micron large, making it possible to resolve the microstructure with an optical microscope~\cite{Bremer1989,Lucey1997, Leocmach2015}.

A common method for inducing homogeneous acidification and thus casein gelation is to introduce glucono-$\delta$-lactone (GDL) to the casein solution~\cite{Bremer1989, Roefs1990, Lucey1997, Lucey1998, Leocmach2014, Leocmach2015, keshavarzNonlinearViscoelasticityGeneralized2017}. GDL hydrolyses into gluconic acid with a slow kinetic, lowering the pH to a final value that should be close to the isoelectric point of casein (4.6) in order to have a stable gel. The initial quantity of GDL sets both the final pH and the delay between mixing and gelation. Typical gelation times are larger than 8 hours. Introducing more GDL induces faster gelation, to the cost of unstable gels that swell and partially redissolve as the pH falls below 3.5~\cite{Leocmach2015}.

In ICAMM, the gel is immersed in its own solvent in order to avoid surface tension effects. The surrounding solution should be slightly lighter than the gel-forming solution, otherwise the latter would not stay at the bottom. Furthermore, the pH of the surrounding solution cannot be constant and should be close to the isoelectric point, otherwise gelation would occur heterogeneously at the contact between the gel-forming solution and the surrounding solution. Therefore, the pH of the surrounding solution should decrease with time, reach the gelation pH after the gel-forming solution does, and stabilize at a pH close to the isoelectic point. As mentioned above, GDL could lower the pH in this way in about 8 hours. To speed up this process, we use the buffering power of acetate ($\mathrm{pK}_{a} = 4.75$), able to stabilize pH close to the isoelectric point of casein.

In practice, we prepare an acetate solution by mixing \SI{0.764}{\% wt} sodium acetate-trihydrate (\ce{CH3COONa}.3\ce{H_2O}, Sigma-Aldrich CAS: 6131-90-4) and  $\SI{0.320}{\% v}$ of glacial acetic acid (\ce{CH3COOH}, VWR Chemicals CAS 64-19-7) in $9:1$ ratio so that the molar ratio in the final solution is $[\ce{CH3COOH}]/[\ce{CH3COO-}] = 0.1$. This solution has a pH $\approx 5.75$. Upon addition of \SI{0.75}{\% wt} GDL (TCI CAS: 90-80-2), the pH decreases to 5.2 in \SI{35}{\minute}, and then slowly converges to its equilibrium pH $\approx 4.4$, as shown in Fig.~\ref{fig:pH_time}. 

To prepare the gel-forming solution, we dissolve \SI{1}{\% wt} sodium caseinate (TCI CAS: 9005-46-3) in water at room temperature and mix this solution in $1:1$ volume ratio with an acetate solution made by mixing  \SI{1.528}{\% wt} sodium acetate-trihydrate and \SI{0.320}{\% v} of acetic acid in $9:1$ ratio. The concentration in acetate is thus similar between the gel-forming and the surrounding solution. The initial pH of the gel-forming solution is 5.9. Before the experimentation and addition of GDL, we add Rhodamine B dye (Sigma Aldrich CAS: 81-88-9)  so that we have a concentration of \SI{2}{\micro \mole} in the gel forming solution. Upon addition of \SI{2.5}{\% wt} GDL, the pH decreases to 5.2 in 20 min and then converges to its equilibrium pH $\approx 3.6$, as shown in Fig.~\ref{fig:pH_time}. 

To prepare an experiment in ICAMM, we start from an empty tank. The head is pressed onto the bottom coverslip by a physical contact between the sensors and their target. Then, we add GDL simultaneously to both the gel-forming and the surrounding solutions. After \SI{10}{\second} mixing, we immediately pipette \SI{200}{\micro\litre} of the gel-forming solution around the head. Then we fill the tank with 50-\SI{60}{\milli\litre} of surrounding solution. Part of the filling is done at a controlled flow rate of \SI{60}{\milli\litre/\hour}, using a syringe pump and a \SI{0.3}{\milli\metre} inner diameter tube ending at the end of the tank close to the head, in order to minimize the mixing with the gel forming solution. Once the perimeter of the head is surrounded, the filling is completed manually with a pipette from the other end of the tank. When the tank is filled, we raise the head by \SI{100}{\micro\metre} and wait 40-\SI{45}{\minute} for the gelation to take place after which, we put a control loop for \SI{135}{\minute} before any test to ensure all chemical species are in equilibrium between the surrounding and the gel. Mixing of the two solutions does occur before gelation, especially at the beginning of the filling of the tank. However, the gel-forming solution is denser and stays in the hollow around the head. Furthermore, the tight confinement by the touching sphere-plane geometry prevents mixing under the head itself. When the head is raised, the composition in the gap far from its edge is the one of the pure gel-forming solution. Indeed, gelation is observed \SI{20}{\minute} after mixing as in the pure gel-forming solution. The typical diffusion time between the axis of the head and its edges is $r_c^2/(2D_{\ce{H+}}) \approx \SI{32}{\minute}$ for \ce{H+} ions. Therefore, we consider that one hour after mixing, the pH is set by the pH of the surrounding solution, slowly decreasing between 4.9 and 4.4, which is close enough to the isoelectric point of casein to have a stable gel.

We use the same gel-forming and surrounding solutions in the plane-plane geometry (rotor diameter \SI{4.3}{\centi\metre}, gap size \SI{1}{\milli\metre}) of a rheometer (Anton Paar MCR 301). The sample is injected in the gap and surrounded with the buffer. The peak value of the elastic modulus measured (\SI{1}{\%} strain, \SI{1}{\hertz}) is $G^\prime = \SI{7.05 \pm 0.25}{\pascal}$.

\subsection{PID control}

\begin{figure}[t]
\centering
\includegraphics[width=\columnwidth]{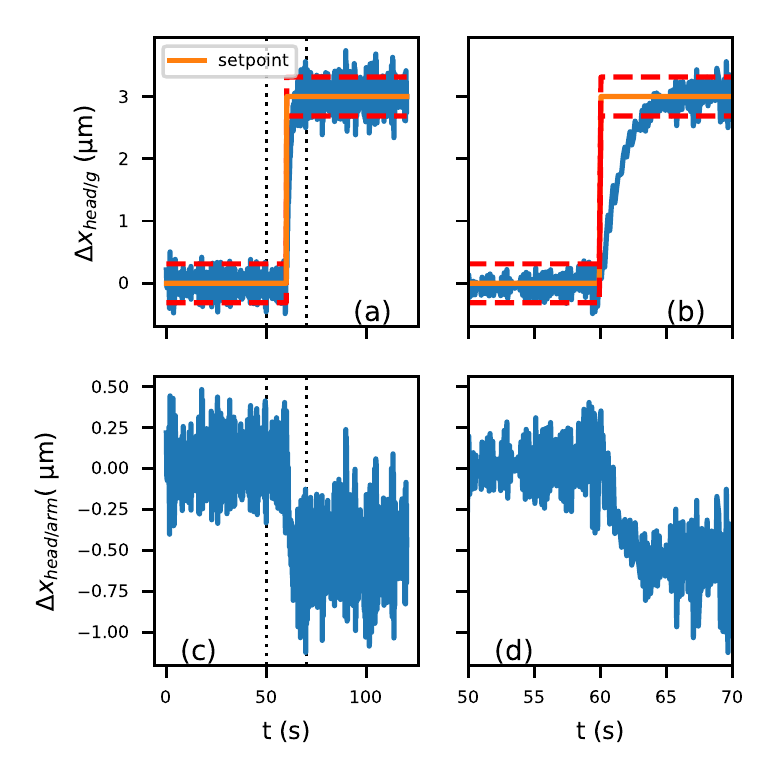}
\caption{Step strain response: (a) The desired set-point (in orange) for a step displacement in $x$ of \SI{3}{\micro\metre} and the actual displacement response (blue) with time. The red dashed lines show the steady-state error of the proportional controller $\pm e_p$ with $K_{p} = 0.1$. During the entire duration, we keep the strain in $z$ constant. (b) Zoom  $\pm\SI{10}{\second}$ (dotted lines in (a)) before and after the transition in displacement $\Delta x_{head/ground}$. (c) change in deflection $\Delta x_{head/arm}$ corresponding to the applied strain and (d) zoomed of $\Delta x_{head/arm}$ around the transition time}.
\label{fig:step_strain}
\end{figure}

Mechanically, ICAMM is neither a stress-controlled or strain-controlled setup. Indeed in most practical cases, the stiffness of the cantilever is close to the equivalent stiffness of the studied sample. For example, a gel with $G^\prime \approx \SI{7}{\pascal}$ in our geometry has an equivalent stiffness of $k_\mathrm{gel} = G^\prime A_\mathrm{eff}/h_0 \approx \SI{2}{\newton/\metre}$, similar to the stiffness of the cantilever. That is why we need to introduce a feedback control on either the position of the head with respect to the ground or the deflection in order to obtain a strain-controlled, respectively stress-controlled, test. As shown in Fig.~\ref{fig:pid}, we can set this mode on both axis independently. In the following, we will perform only shear tests in the $x$ direction, maintaining a constant gap thickness with a strain-control on $z_\mathrm{head/ground}$.

The PID controller acts by calculating the error $e(t)$, which is the difference between the set point and the process variable, and acting on this $e(t)$ using a proportional (P), Integral (I), derivative (D) correction so as to minimize the quantity
\begin{equation}
   PV =K_{p}e(t) + K_{i}\int_{0}^{t} e(t^{'})d(t^{'}) + K_{d}\frac{de(t)}{dt} 
\end{equation}
where $K_p$, $K_i$ and $K_d$ are coefficient of the P, I and D control respectively.

Since the micromanipulator moves in steps of finite size ($\epsilon =\SI{62,5}{\nano\metre}$), a purely proportional controller ($K_p>0$, $K_i=K_d=0$) cannot correct an error such that $|e(t)|< e_p$, where $e_p = \epsilon/(2 K_p)$ is the steady-state error of the proportional controller. This error can be improved by using a larger value of $K_p$ or by introducing an integral controller which keeps adding the error over time. Either of these action can lead to overshoot and instability in control loop and hence, a further differential controller can be added,  which anticipates the rate of change in $e(t)$ and dampens it. Also, the frequency of our control loop is limited by the frequency of action of micromanipulator which is $\SI{10}{\hertz}$.

\subsection{Step strain}

To test our control loop, we do a step strain experiment and record the stress response from the deflection of the cantilever. 
As shown in Fig.~\ref{fig:step_strain}a, we fix a set point at $x_\mathrm{head/ground}=\SI{0}{\micro\metre}$ for \SI{60}{\second}, and then update the set point to $x_\mathrm{head/ground}=\SI{3}{\micro\metre}$. This corresponds to a shear strain of $0.03$. The PID ($K_{p} = 0.1$, $K_i=K_d=0$) controller acts on $x_\mathrm{head/ground}$, that is to say the shear stain of the gel.

In Fig~\ref{fig:step_strain}b, we see the zoom of Fig~\ref{fig:step_strain}a $\pm \SI{10}{\second}$ around the update of the set point. The controlled variable $x_\mathrm{head/ground}$ converges to the set point in \SI{5}{\second}. To speed up the response, we can increase the $K_p$ or use a PI controller. Fig~\ref{fig:step_strain}c shows the change in deflection $\Delta x_{head/arm}$, a measure of shear-stress. The deflection  $\approx 0.5\mu$m corresponds here to a shear force $\approx 1\mu$N. Fig~\ref{fig:step_strain}d is the zoom of Fig~\ref{fig:step_strain}c on same time scale as Fig~\ref{fig:step_strain}b. We see clearly a progressive stress shift at the transition confirming that the gel is attached and responding to the head motion.

\begin{figure}
\centering
\includegraphics[width=\columnwidth]{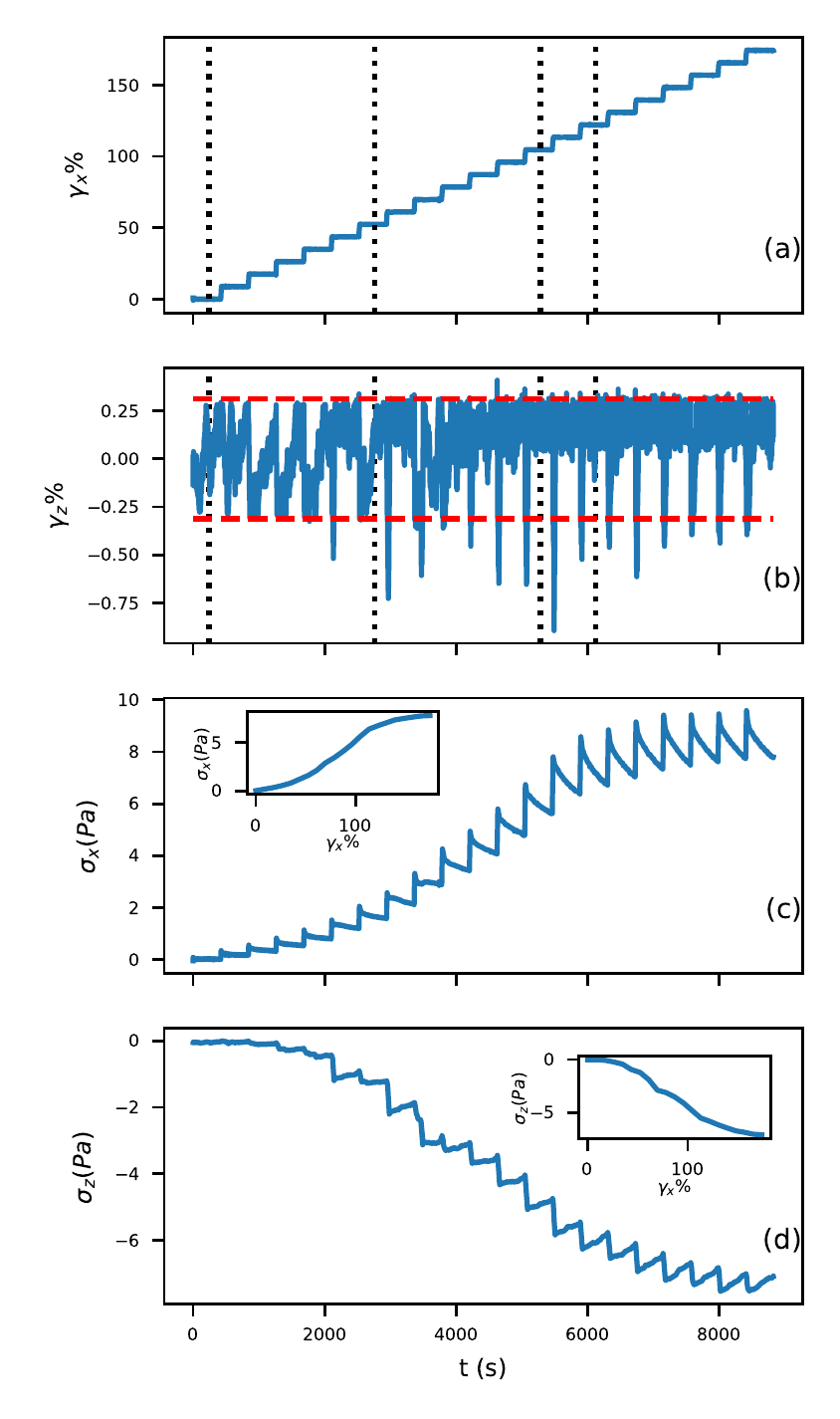}
\caption{Behavior at large strain: shear strain setpoint increasing in steps of \SI{10}{\%} every \SI{420}{\second} from $0 \%$ to $200 \%$  for a constant normal strain setpoint. (a) Measured shear stain. (b) Measured normal strain. The red dotted line indicate the steady-state error of the proportional controller (c) The measured shear stress and in inset the shear stress averaged over the last $\SI{10}{\second}$ of the step as a function of shear strain and (d) normal stress variation with time and in inset the normal stress averaged over the last $\SI{10}{\second}$ of the step vs shear strain. Dotted vertical lines in (a) and (b) mark the times of the pictures in Fig.~\ref{fig:confocal}.}
\label{fig:strain_stairs}
\end{figure}

\subsection{Shear strain steps and simultaneous confocal acquisition}
\label{strain_steps_confocal}

We can repeat strain steps to test the mechanical behaviour of the gel at larger strains. Here, we start the test \SI{210}{\minute} after mixing of GDL. In Fig.~\ref{fig:strain_stairs}a, we show the strain $\gamma_x$ applied using a proportional controller ($K_{p} = 0.1$, $K_i=K_d=0$), in  which set point for $x$ position increases by steps of \SI{10}{\micro\metre} (\SI{8.6}{\%}) strain every \SI{7}{\minute} until a shear strain of \SI{172}{\%} (set point not shown). The gap is kept constant at $h_0=\SI{115\pm0.3}{\micro\metre}$ by a second proportional controller with same constant, leading to normal strain fluctuations $\delta\gamma_y \approx 0.25\%$ as shown in Fig.~\ref{fig:strain_stairs}b. After each step, the shear stress shown in Fig.~\ref{fig:strain_stairs}c displays the same non-linear viscoelastic relaxation as reported in Ref.~\cite{keshavarzNonlinearViscoelasticityGeneralized2017}. By averaging the last \SI{10}{\second} of each step, we obtain the stress-strain dependence (inset of Fig.~\ref{fig:strain_stairs}c). Overall, the gel is strain hardening between 34~\% and 121~\%, and strain softening at larger strains. Compared to Ref.~\cite{keshavarzNonlinearViscoelasticityGeneralized2017}, our gel displays a much larger strain hardening domain, due to the four time lower casein concentration.
From the linear regime at strain below 34~\%, we extract an elastic modulus $G^{\prime}=\SI{1.986 \pm0.085}{\pascal}$.  The error includes the systematic uncertainty (see section~\ref{system_uncertainity}). Taking into  account the $G\propto \omega^0.15$ scaling of casein gels~\cite{Leocmach2014} and the low equivalent frequency of our measurements ($\approx 1/\SI{7}{\minute}$) the rheometer measurement at \SI{1}{\hertz} (see Section~\ref{sec:sample_prep}) interpolates to $\approx\SI{2.9}{Pa}$. The lower value measured by strain steps can be attributed to the difference in gelation procedures, normal force conditions~\cite{maoNormalForceControlled2016b,pommellaCouplingSpaceResolvedDynamic2019} and the difference in the rheological procedures. Fig.~\ref{fig:strain_stairs}d shows that the normal stress is also reliably measured, and follows the shear stress.

\begin{figure*}
\centering
\includegraphics[width=\textwidth]{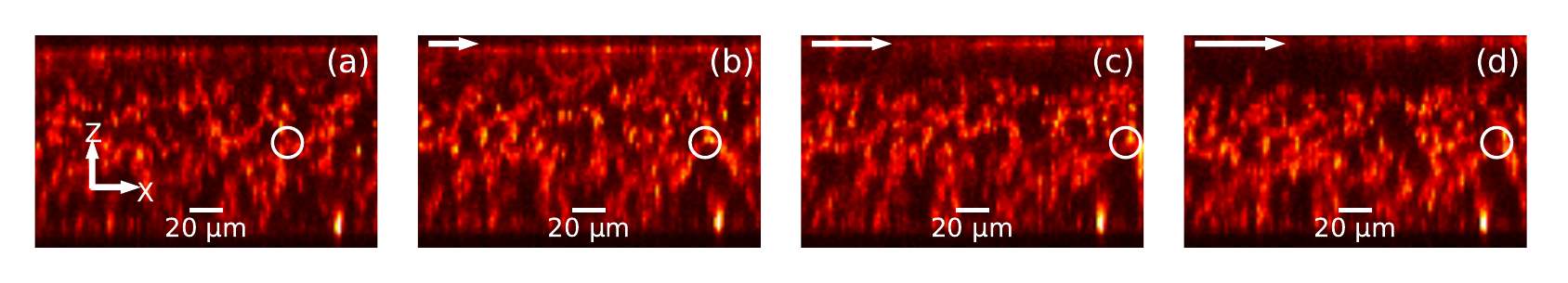}
\caption{Microstructure visualisation during strain steps. From (a) to (d): Cut in the shear plane (x,z) from 3D confocal acquisition at $0 \%$, $52 \%$, $104 \%$ and $121 \%$ of strain respectively. The top of the panel is the head and the bottom the coverslip. The magnitude and direction of shear strain applied to the head is indicated by the white arrows at the top of each panel and by the dotted vertical lines in Fig.~\ref{fig:strain_stairs}. Displayed images are the average of 10 (x,z) planes of the confocal stack for a total thickness of \SI{4.45}{\micro \metre}. The white circle tracks a section of the gel and its approximate trajectory.}
\label{fig:confocal}
\end{figure*}

As a proof of concept, we performed three dimensional confocal acquisition (Leica SP5, \SI{488}{\nano\metre} excitation).
Crucially, we use here an objective lens without immersion fluid (Leica HC PL APO $40\times$ NA=0.95 CORR). Previous attempts with oil or water immersed objectives have revealed that the immersion fluid was transmitting enough force from the z-scanning objective to bend the cover slip by a few micrometers and perturb the mechanical measurement. Without optical immersion fluid, there is no signature of the z-scanning cycle on the measurements shown in Fig.~\ref{fig:strain_stairs}. We calibrated the pixel-to-micron ratio along the z axis using a cell of known thickness (\SI{113}{\micro\metre}) filled with the gel-forming solution.

We start each stack at the 4$^\text{th}$ minute of each step. The full $228\times 228 \times \SI{139}{\micro\metre}$ stack is acquired in \SI{120}{\second} and is centered on the axis of the sphere-plane geometry. In this way, we obtain a 3D stroboscopic view of the microstructure responding to shear. In Fig.~\ref{fig:confocal}, we show a cut through the acquired volume in the shear $(x,z)$ plane at four different steps: $\gamma=0\%$, $51\%$, $102\%$ and $121\%$. In Fig.~\ref{fig:confocal}a, we qualitatively observe that the density in protein is not constant along z: there is an adsorbed layer on both the cover slip at the bottom and on the head at the top. Furthermore a few microns below the head, the density seems to be lower than in the bulk of the gel. Between Fig.~\ref{fig:confocal}a and c, we observe the progressive shear of the gel network. At $\gamma=111\%$, the adsorbed layer on the head is completely detached from the bulk of the gel in the observable zone and as seen in Fig.~\ref{fig:confocal}d, the bulk of the gel undergoes viscoelastic recoil. Since we do not observe a corresponding drop in the macroscopic shear stress response expected with complete failure (see Fig.~\ref{fig:strain_stairs}c), we think that the fracture does not reach the edge of the head. Indeed, as detailed in Sec.~\ref{sec:area}, further away from the axis of the geometry the gap is larger and the strain and stress are smaller, so that it may not be enough for the fracture to propagate.

 
\begin{figure}
\centering
\includegraphics[width=\columnwidth]{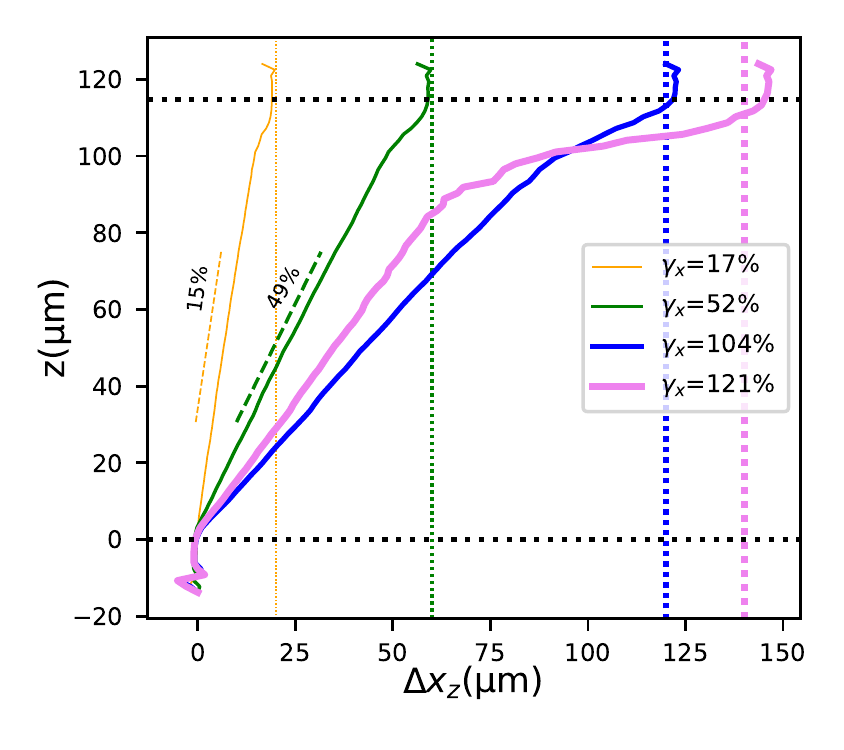}
\caption{Displacement profile in the gap. The displacement in the shear direction ($x$) is obtained by accumulating the image correlation computed value between two consecutive time frames. The four curves are for macroscopic strains $17\%$, $52\%$, $104\%$ and $121\%$. The dotted line next to the curve with slope value correspond to the best fit for the strain in the bulk of the gel. The horizontal dotted line denotes the coverslip (bottom) and the head (top) position. Vertical dotted lines denote the macroscopic imposed strain value for the four plotted curves}
\label{fig:cum_shift_x}
\end{figure}

From the confocal images, we can obtain the displacement profile in the gel ( Fig.~\ref{fig:cum_shift_x}). We use plane by plane 2D image phase correlation between consecutive stacks, and accumulate these displacements from 0 to $\gamma_x$ to obtain a displacement profile $\Delta x(z)$ at each step $\gamma_x$. Since the scanning direction and the shear direction are well aligned, displacements along $y$ are at least two orders of magnitude smaller than along $x$ and are neglected here.

For small strains (see e.g. the yellow curve in Fig.~\ref{fig:cum_shift_x}), we observe that the strain is almost homogeneous in the whole gap, with a linear $\Delta x(z)$ for $\SI{5}{\micro\metre}\leq z\leq\SI{100}{\micro\metre}$. However, we notice that close to the coverslip or the head, the slope is steeper for a few microns, indicating harder materials that corresponds to the adsorbed layers. Furthermore, we observe a smaller slope, i.e. a softer layer, below the head for $\SI{100}{\micro\metre}\leq z\leq\SI{110}{\micro\metre}$. This behaviour is conserved until $\gamma_x=104\%$, with a softening of the already soft layers, probably due to damage accumulation. Finally, at $\gamma_x=121\%$ (pink curve on Fig.~\ref{fig:cum_shift_x}) we observe a complete rupture of the soft layer, where the top layer remains adsorbed on the head. By contrast, the bulk of the gel recoils viscoelastically and also compresses downwards, which widens the fracture and reduces the extent of the linear zone.

This quantitative, space-resolved analysis is a proof of concept, showing that ICAMM can be integrated with confocal microscopy and yield more detailed information than what is capture but the global mechanical response alone.


\subsection{Controlled stress}

\begin{figure}
\centering
\includegraphics[width=\columnwidth]{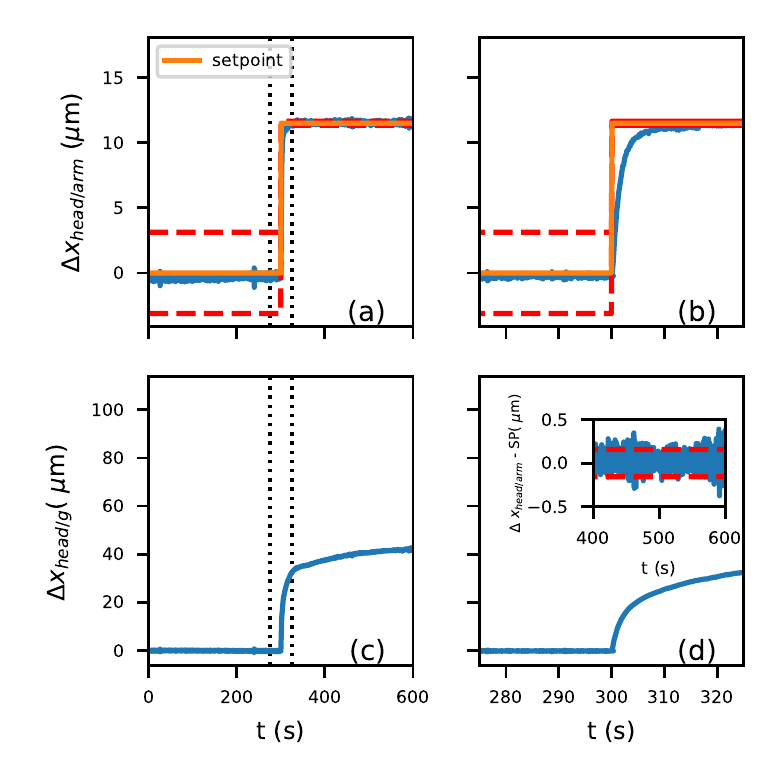}
\caption{Step stress response: (a) The desired set-point (in orange) for a step deflection in $x$ of \SI{11.50}{\micro\metre} and the actual deflection (blue) with time. The red dashed lines show the steady-state error of the proportional controller $\pm e_p$. During the entire duration, we keep the strain in $z$ constant. (b) Zoom $\pm \SI{25}{\second}$  before and after the step in set point, see dotted lines in (a). (c) Change in position $\Delta x_{head/ground}$ corresponding to the applied stress and (d) zoomed-in around the transition time. Inset: Stability of the deflection around the set point at later times.}.
\label{fig:step_stress}
\end{figure}

By controlling the cantilever deflection, ICAMM can also perform step stress experiments and record the strain response. 
Here, the control on $x_\mathrm{head/arm}$ is ensured by a proportional-integral (PI) controller. We use a Ziegler-Nichols method~\cite{zieglerOptimumSettingsAutomatic1993} to optimize the constants of the controller: $K_{p} = 0.45 K_{u}$,
    $K_{i} = 0.54K_u/T_u$,
where $K_u$ is the ultimate proportional gain at which the output displays stable oscillations and $T_u$ is the time period of these oscillation. Since oscillations of diverging amplitude quickly destroy the gel, requiring a new sample each time, we limited ourselves to a range $0.35<K_u<0.5$ and $T_u \approx \SI{40}{\second}$. Exploring from these values, we obtain a stable response without overshoot for $K_p = 0.2$ and $K_i = 0.001$.

As shown in Fig.~\ref{fig:step_stress}a, we fix a set point at $x_\mathrm{head/arm}=\SI{0}{\micro\metre}$ for \SI{300}{\second} with a compliant PI controller ($K_p = 0.01$ and $K_i = 0.001$), and then update the set point to $x_\mathrm{head/arm}=\SI{11.50}{\micro\metre}$, that is to say an effective stress $\sigma_0=\SI{0.79}{\pascal}$, with the tighter controller determined above ($K_p = 0.2$ and $K_i = 0.001$). We expect a steady state error of \SI{0.02}{\pascal}, further narrowed by the integral term with a time constant of the order of \SI{20}{\second}.

In Fig~\ref{fig:step_stress}b, we see the zoomed in version of Fig~\ref{fig:step_stress}a to $\pm \SI{25}{\second}$ around the update of the set point. The controlled variable $x_\mathrm{head/arm}$ converges to the set point in \SI{10}{\second} and remains stable on much longer times (inset of Fig~\ref{fig:step_stress}d). Fig~\ref{fig:step_stress}c shows the change in position $\Delta x_{head/ground}$, a measure of shear-strain. Fig~\ref{fig:step_stress}d is the zoomed in version of Fig~\ref{fig:step_stress}c on same time scale as Fig~\ref{fig:step_stress}d. We see clearly that the strain evolves in time even after the stress has settled to its set point value.

\begin{figure}
\centering
\includegraphics[width=\columnwidth]{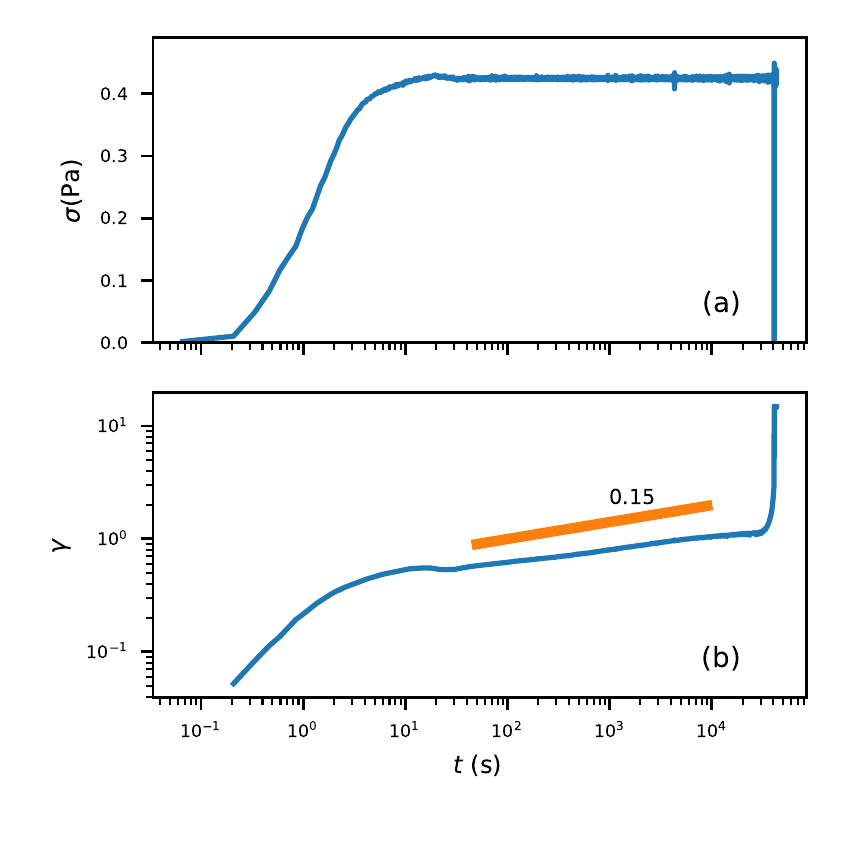}
\caption{Constant shear stress experiment: at $t=0$, the set point jumps from zero to $\sigma=\SI{2.13}{\pascal}$ for a gel where $G^{\prime}=\SI{2.13}{\pascal}$ The gap is kept constant at $h_o =  \SI{103.8}{\micro \metre}$.
(a) The actually applied stress function of time. The set point is reached in \SI{10}{\second} without overshoot.
(b) The shear strain response function of time in log-log scale. The straight line highlights the power-law regime after initial convergence of the feedback loop.
}
\label{fig:creep}
\end{figure}

\subsection{Creep experiment}

Finally, we demonstrate that the ICAMM is able to study the long time response to constant shear stress. For the procedure, after mixing GDL, we keep the gel under no control for the initial 45 min and then apply zero force in both shear and normal direction for the next $\approx \SI{135}{min}$ ($K_{p} = 0.01, K_{i} = 0.0005$). Then, we estimate the elastic modulus by performing small strain steps of $3\%$ from $0\%$ to $9\%$ in step,  $\SI{60}{\second}$ each. A linear fit of the stress response (not shown) gives $G^{\prime}=\SI{1.418\pm0.156}{\pascal}$.
$\SI{210}{\minute}$ after mixing, we change the set point in the $x$ direction to $\sigma=\SI{2.13}{Pa}$, i.e. a deflection of $\SI{31.10}{\micro\metre}$  with $K_{p} = 0.2$ and $K_{i} = 0.001$. In the $y$ direction, the gap is kept constant $h_o =  \SI{103.8}{\micro \metre}$ with $K_{p} = 0.1$ and $K_{i} = 0.001$. The actual applied stress is shown in Fig.~\ref{fig:creep}a. It converges to its set point in $\approx \SI{10}{\second}$. 

In Fig.~\ref{fig:creep}b, we show the evolution of the strain in log-log scale. We clearly observe at short times ($<\SI{10}{\second}$) the regime where the response is dominated by the convergence of the feedback loop. At intermediate time scale, the stress is properly applied and can be considered constant. We observe the power-law regime characteristic of the frequency-dependent viscoelastic response of casein gels $\gamma\sim t^\alpha$~\cite{Leocmach2014}   with a similar value of the exponent $\alpha \approx 0.15$. At later times, we observe the divergence of the strain that indicates nucleation and growth of fractures. Finally, the gel undergoes full rupture.

\section{Conclusion}
We have developed a robust setup to probe the long term mechanical response of soft materials to steady stimuli while having a direct microscopic visualization of the structural change happening inside them. The large dynamic range of the sensors can help explore materials ranging from very soft (\SI{10}{\milli\pascal}) to soft (\SI{10}{\pascal}). ICAMM can control either stress or strain independently in shear and normal direction. 

The drawback of our design is the long $\approx \SI{10}{\second}$ response time of ICAMM. This makes our apparatus unsuited for steady shear-rate experiments or for oscillatory rheology at high frequencies. The response time could be shortened by using a different actuator with faster electronics. However the inertia of the cantilever and viscous forces acting on the head would set a lower bound for the response time.

The most promising aspect of ICAMM is its ability to obtain direct visualization of the microstructure of soft materials under well-controlled steady mechanical stimuli. We have demonstrated the use of plane-by-plane image correlation to obtain the displacement profiles during controlled strain experiment. This could be extended to other kind of experiments and refined to obtained more local strain field. In particular, we intend to use ICAMM to understand the microscopic origin for macroscopic rheology behavior in case of phenomena such as creep and yield in soft solids.

By reducing the radius of curvature of the head, one can reduce the effective area to sizes comparable to the field of view of a microscope. This would bring into view all relevant fracture precursors, enabling the study of fracture nucleation. Additionally, increasing the length or decreasing the cross-section of the cantilever would provide even higher sensitivities. This would enable to reliably apply stress to extremely soft gels made of micron-size colloidal particles. In these systems, one could study at single-particle level the diffuse damage that precedes fracture nucleation.


\begin{acknowledgments}
We want to thanks Hélène Delanoë-Ayari, Catherine Barentin and Loren J\o{}rgensen for drawing our attention onto the tension and compression cantilever setup and sharing with us the improvement possibilities. We acknowledges support from ANR grant GelBreak ANR-17-CE08-0026. 
M. T. and M. L. acknowledge funding from CNRS through PICS No. 7464. Initial funding has been provided by FRAMA and Institut de Chimie de Lyon.
\end{acknowledgments}

\bibliography{biblio.bib}

\end{document}